# Article

# Sympathetic cooling of a trapped proton mediated by an LC circuit

M. Bohman[1,2 ✉], V. Grunhofer[3], C. Smorra[2,3], M. Wiesinger[1,2], C. Will[1], M. J. Borchert[2,4,5], J. A. Devlin[2,6], S. Erlewein[2,6], M. Fleck[2,7], S. Gavranovic[3], J. Harrington[1,2], B. Latacz[2], A. Mooser[1], D. Popper[3], E. Wursten[2,6], K. Blaum[1], Y. Matsuda[7], C. Ospelkaus[4,5], W. Quint[8], J. Walz[3,9], S. Ulmer[2] & BASE Collaboration*

Efficient cooling of trapped charged particles is essential to many fundamental physics experiments[1,2], to high-precision metrology[3,4] and to quantum technology[5,6]. Until now, sympathetic cooling has required close-range Coulomb interactions[7,8], but there has been a sustained desire to bring laser-cooling techniques to particles in macroscopically separated traps[5,9,10], extending quantum control techniques to previously inaccessible particles such as highly charged ions, molecular ions and antimatter. Here we demonstrate sympathetic cooling of a single proton using laser-cooled Be⁺ ions in spatially separated Penning traps. The traps are connected by a superconducting LC circuit that enables energy exchange over a distance of 9 cm. We also demonstrate the cooling of a resonant mode of a macroscopic LC circuit with laser-cooled ions and sympathetic cooling of an individually trapped proton, reaching temperatures far below the environmental temperature. Notably, as this technique uses only image–current interactions, it can be easily applied to an experiment with antiprotons[1], facilitating improved precision in matter–antimatter comparisons[11] and dark matter searches[12,13].

Measurements of the charge-to-mass ratio and $g$-factor of the proton and antiproton, a prominent, stable particle–antiparticle system, are limited by cryogenic particle temperatures[14–16]. However, with no electronic structure, (anti)protons (protons and antiprotons) are not amenable to standard laser cooling techniques. Moreover, ions that are easily laser cooled are not readily trapped in the same potential well as negatively charged antiprotons or antimatter molecular ions (for example, $\overline{H_2}$)[17]. Sympathetic laser cooling with negatively charged ions[18–20] and with microscopically fabricated trapping potentials[1,21,22] have been proposed. Another technique, proposed over 30 years ago, extended laser cooling to exotic systems by coupling via induced image currents in trap electrodes to ions with a well-suited cooling transition[9]. Similarly, coupling laser addressable ions to systems with no optical structure is sought after in precision spectroscopy[23,24], mass measurements[25], quantum information[10] and quantum engineering[5].

We demonstrate sympathetic cooling of a single proton, extending the image-current coupling technique with a superconducting LC circuit that resonantly enhances energy exchange between the proton and laser-cooled ions. We use a cryogenic multi-Penning-trap system to store a single proton in the proton trap and a cloud of Be⁺ ions in a beryllium trap, separated axially by around 9 cm (Fig. 1a). A homogeneous magnetic field $B$ parallel to the electrode axis and an electric quadrupole potential at voltage $V_0$ confines the particles and gives rise to circular magnetron and modified cyclotron motion in the radial plane, and harmonic axial motion, at frequencies $v_-$, $v_+$ and $v_z$, respectively[26]. LC resonators with high quality factors (in our case $Q \sim 15{,}000$) are commonly used to detect image currents in trap electrodes[27] and, as shown in Fig. 1a, we connect the resonator to both traps so that the two ion-trap systems are coupled via particle-induced image currents. The LC circuit with total capacitance $C_R \approx 36$ pF and inductance $L_R \approx 3.0$ mH has an equivalent parallel resistance, at resonance frequency $v_0$, of $R_p = 2\pi v_0 L_R Q$. The entire system is modelled by an equivalent circuit where the proton and Be⁺ ions are series LC circuits with capacitances and inductances $C_p$, $C_{Be}$, $L_p$ and $L_{Be}$ (ref. [28]) connected in parallel to the superconducting LC circuit (Fig. 1b). The Be⁺ ions are also damped by the cooling laser, represented as a variable virtual resistance $R_L$ (ref. [9]). In contrast to the proposal in which energy is exchanged between the ion-trap systems via only a shared electrode[9], we use the LC-circuit resonator to couple the axial modes of the trapped particles. On resonance, the large inductance of the resonator coil compensates the electrode capacitance and enhances the ion-induced image current by the $Q$-value. With mechanically machined traps used for precision Penning-trap experiments, the ~10-mHz coupling rates expected from the non-resonant proposal[11] require minute-scale cooling cycles and are limited by the loss of resonant coupling; for example, from voltage fluctuations of the trapping potential. For the parameters used in our resonant cooling demonstration, energy is exchanged between the proton and sympathetically laser-cooled resonator at a rate of 2.6 Hz (measured by the dip width on resonance) so that thermal equilibrium is reached within seconds and the axial frequencies of the two species are easily matched. Notably, by coupling the ion-trap systems via the resonator, the coupling does not rely on a shared electrode, so the energy exchange rate is not limited by the trap capacitance. Consequently, this cooling scheme can be realised over long distances and with several distributed ion traps.

[1]Max-Planck-Institut für Kernphysik, Heidelberg, Germany. [2]RIKEN, Fundamental Symmetries Laboratory, Saitama, Japan. [3]Institut für Physik, Johannes Gutenberg-Universität, Mainz, Germany. [4]Institut für Quantenoptik, Leibniz Universität Hannover, Hannover, Germany. [5]Physikalisch-Technische Bundesanstalt, Braunschweig, Germany. [6]CERN, Geneva, Switzerland. [7]Graduate School of Arts and Sciences, University of Tokyo, Tokyo, Japan. [8]GSI Helmholtzzentrum für Schwerionenforschung GmbH, Darmstadt, Germany. [9]Helmholtz-Institut Mainz, Mainz, Germany. *A list of authors appears at the end of the paper. ✉e-mail: matthew.bohman@mpi-hd.mpg.de

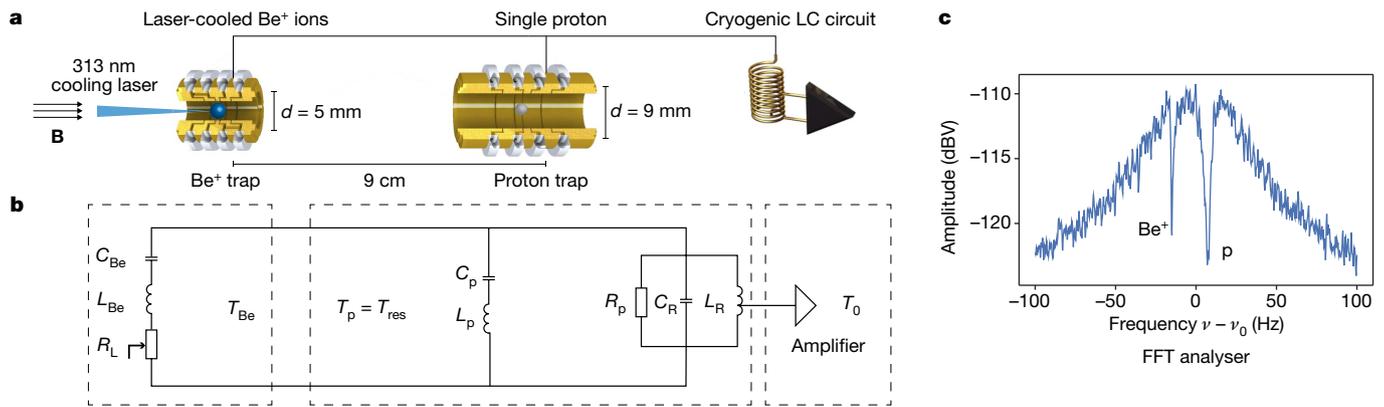

**Fig. 1 | Experimental setup. a**, A single proton is stored in the proton trap while one or more Be⁺ ions are stored in the beryllium trap. The two ion traps, with inner diameters $d = 9$ mm and $d = 5$ mm, respectively, are connected to a cryogenic LC circuit with resonance frequency near their axial frequencies. One end of the resonance circuit is connected to a cryogenic amplifier while the other is connected to rf ground. **b**, This system is described by a three-part equivalent circuit, two series LC circuits representing the trapped particles, and the LC resonator with effective parallel resistance $R_p$ as a parallel RLC circuit. The amplifier is used to read out the image–current signal of the circuit, and drives the system with voltage noise at an effective temperature $T_0$. **c**, The resulting FFT spectrum consists of the broad resonance of the detector and two narrow 'dips' with ~0.8 Hz FWHM for a single Be⁺ ion and ~2.6 Hz FWHM for the proton at the axial oscillation frequencies of the two trapped-ion systems.

The noise spectrum of this coupled system is shown in a fast Fourier transform (FFT) of the voltage signal of the resonator (Fig. 1c). The entire system is driven by a combination of the Johnson noise of the resonator and additional voltage noise from the cryogenic amplifier, resulting in an effective noise temperature $T_0 = 17.0(2.4)$ K (where the number in parentheses is the 1$\sigma$ uncertainty). Here, the axial frequencies of the proton in the proton trap and the Be⁺ ion in the beryllium trap were set close to resonance with the LC circuit by adjusting the voltage of the axial potential in each trap, $v_z \propto V_0^{1/2}$. In the measured noise spectrum, the detector appears as a broad ~40 Hz (full-width at half-maximum (FWHM)) resonance while the proton and the Be⁺ ions short the parallel resistance of the resonator and appear as narrow dips with widths determined by the charge-to-mass ratios and the trap diameter[28].

We demonstrate that the proton, Be⁺ ions and resonator form a system of three coupled oscillators by measuring the noise spectrum at thermal equilibrium. We detune both ion species around three resonator linewidths away from the LC-circuit resonance frequency $v_0 \approx 479{,}000$ Hz to observe coupling signatures via the FFT lineshape. In these measurements, we store the proton in the proton trap at constant axial frequency and gradually increase the axial frequency of a single Be⁺ ion in the beryllium trap. The resulting FFT spectra show observed particle frequencies at the dip positions (Fig. 2a, dark blue) and two of the normal modes of the coupled three-oscillator system at the maxima (Fig. 2a, red). Near the p-Be⁺ resonance, the axial motion of both particles is no longer determined by the trapping potential alone, and we observe the coupling signature in two of the normal modes of the three-oscillator system. This feature is consistent with the analytical solution derived from the impedance of the circuit model in Fig. 2b (see Methods) and appears when the three oscillators exchange energy. Using simulations (described in the Methods), we show the corresponding time-domain behaviour in the presence of the environmental noise (Fig. 2c). In the absence of environmental noise, the energy of each oscillator as a function of time is deterministic and can be found from the initial phases and energy exchange rates. With environmental noise included, energy is still exchanged and, as shown here when laser cooling is absent, the oscillator energies are determined by this equivalent noise temperature.

We further demonstrate that the temperature of the proton can be modified by coupling to a cloud of excited Be⁺ ions, here consisting of around 15 ions. To this end, we apply a parametric rf drive at $2v_0$ in the beryllium trap, which excites the Be⁺ ions if $v_{z,\text{Be}} = v_0$ but, as confirmed by background measurements (see Methods), has no direct effect on the proton in the proton trap. By bringing the proton into resonance with the weakly excited Be⁺ ions the Be⁺ ions appear as a broad, shallow dip and the sympathetically excited proton appears as a narrow peak (see Methods). To quantify the energy transferred to the proton, we measure the axial frequency of the proton before and after coupling to the excited Be⁺ ions. Coupling the excited axial mode to the cyclotron mode with a sideband drive at $v_+ - v_z$ transfers the energy of the axial mode to the cyclotron mode with resulting energy $E_+ = (v_+/v_z)E_z$ (ref. [29]). Similar to the continuous Stern–Gerlach effect[30], the quadratic component of the magnetic field in the proton trap, $B_2 = -0.39(11)$ T m⁻², interacts with the magnetic moment of the modified cyclotron mode at energy $E_{+,p}$, producing the axial frequency shift $\Delta v_z \propto B_2 \Delta E_{+,p}$ (see Methods), which we measure to determine the change in axial energy of the proton. We show the evolution of the standard deviation of the change in the proton energy $E_{+,p}$ while the excited Be⁺ ions are tuned to resonance with the proton (Fig. 2d, orange) and when detuned from the resonator (Fig. 2d, blue). During this experiment we interleave on-resonance and off-resonance measurements and see a clear increase in proton energy that arises from the remote, resonator-mediated coupling to the excited ions. In contrast to the off-resonant points, the resonance points exhibit scatter that is nearly three orders of magnitude larger, with a statistical significance of more than 20$\sigma$. The excitation drive remains on during both of the interleaved measurements to ensure that the increased scatter is due only to the ion–proton coupling. In addition, we constrain variations in the resonator temperature due to off-resonant coupling to the excitation drive (see Methods). For comparison, the scatter when the drive is off is also shown in Fig. 2d (green points).

Our demonstration of sympathetic cooling employs similar axial frequency shift measurements in the presence of a continuously laser-cooled Be⁺ ion cloud. The Be⁺ ions are cooled with the closed $^2S_{1/2} \rightarrow {}^2P_{3/2}$ transition and tuned to resonance with the superconducting circuit and the proton (Fig. 3a). The cooling laser damps the axial motion, increasing the equivalent resistance $R_L$ (Fig. 1b) and reducing the signal of the broad Be⁺ dip. The laser-cooled ions reduce the effective noise temperature in the entire circuit and lower the temperature in a narrow frequency range. Using the narrow proton dip as a temperature sensor for the cooled common mode of the system, we determine the temperature reduction experimentally with well understood energy-dependent shifts of the axial dip and develop further insight into the cooling using time-domain simulations.

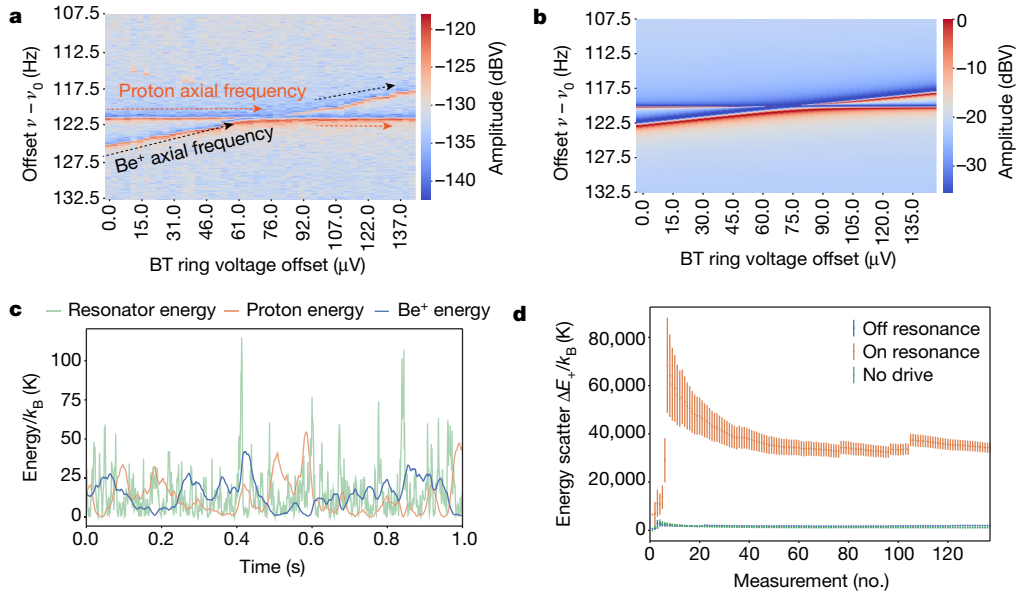

**Fig. 2 | Remote coupling measurements. a**, Measured FFT spectra of a single Be$^+$ ion and the proton are presented as vertical cuts of the heatmap. The axial frequency of the proton is fixed and the ring voltage of the trapping electrodes of the Beryllium trap (BT) is scanned one step upwards after each FFT is recorded. A vertical cross-cut looks similar to the spectrum shown in Fig. 1a, although as the dip features are shifted approximately 130 Hz upwards, they exhibit dispersive behaviour. **b**, The noise spectra expected from the impedance $Z(\nu)$ of the equivalent circuit in Fig. 1b are presented in the same way as in **a** and are calculated using the same parameters as in the experiment. **c**, An example of the energy exchange between the three oscillators at the resonance frequency of the LC circuit is simulated and includes the energy fluctuations from environmental noise of the system. The energy of each oscillator depends on the phase relation with the environmental noise, so the time series shown here is one of infinitely many that could produce the FFT spectrum used in Fig. 2a. **d**, The evolution of the standard deviation of the proton energy scatter is shown, measured via energy dependent axial frequency shifts. We see an unambiguous increase in energy when the excited Be$^+$ ions are tuned to resonance with the proton (orange points), compared to when the Be+ ions are detuned (blue points). The excitation drive remains on for both points and can be compared to the drive off condition (green points). Error bars, $1\sigma$ error of the standard deviation.

A symmetric, cylindrical Penning trap provides a high degree of control over the trapping potential. We use a deliberately introduced trap anharmonicity in the proton trap that shifts the axial frequency by

$$\delta\nu_z(\mathrm{TR}, T) = \frac{1}{4\pi^2 m \nu_z} \frac{3}{2} \frac{C_4(\mathrm{TR})}{C_2(\mathrm{TR})} k_B T. \tag{1}$$

Here, $C_n(\mathrm{TR})$ are the coefficients of the expansion of the local trapping potential along the trap axis that depend on the ratio of voltages applied to the central ring electrode ($V_0$), and the two nearest correction electrodes ($V_\mathrm{CE}$)[26,31,32], referred to as the tuning ratio, $\mathrm{TR} = V_\mathrm{CE}/V_0$. When the laser-cooled Be$^+$ ions are tuned to resonance, the noise energy of the common mode of the proton, resonator and Be$^+$ ions is reduced from the noise temperature of the environment resulting in an axial frequency shift,

$$\Delta\nu_z(T_0, T_p, \Delta\mathrm{TR}) = \nu_{z,1}(\mathrm{TR}, T_0) - \nu_{z,2}(\mathrm{TR}, T_p) \tag{2}$$

$$= \kappa \, \Delta\mathrm{TR} \times (T_0 - T_p), \tag{3}$$

where $\nu_{z,1}(\mathrm{TR}, T_0) = \nu_z + \delta\nu_z(\mathrm{TR}, T_0)$ is the axial frequency measured at $T_0$ when the Be$^+$ ions are detuned and $\nu_{z,2}(\mathrm{TR}, T_p) = \nu_z + \delta\nu_z(\mathrm{TR}, T_p)$ is the axial frequency measured when laser-cooled ions are in resonance and reduce the temperature to $T_p$. The trap anharmonicity is characterized by an offset from the ideal tuning ratio $\Delta\mathrm{TR} = \mathrm{TR} - \mathrm{TR}(C_4 = 0)$ and a constant determined from the trap geometry $\kappa = 45.4$ Hz K$^{-1}$ that we cross-check with additional measurements that use electronic feedback to change the temperature of the resonator. We measure $\Delta\nu_z$ as a function of $\Delta\mathrm{TR}$ and the measured slope $s$ determines the change in temperature,

$\Delta T = T_0 - T_p = -s/\kappa$. The results of an example measurement are shown in blue in Fig. 3b. With ten Be$^+$ ions in resonance, we measure a slope $s = -350(14)$ Hz and in a background measurement with the Be$^+$ ions detuned (Fig. 3b, orange), obtain a slope $s = 4(13)$ Hz. This corresponds to a temperature reduction of $\Delta T = 7.7(0.3)$ K and demonstrates sympathetic laser cooling of a single trapped proton. With a significance of more than 20 standard deviations, this is also a demonstration of remote, image-current mediated sympathetic cooling, applicable to any charged particle without convenient cooling transitions.

The temperature of the proton is determined by the noise power dissipated by the laser-cooled ions. In the circuit representation, increasing the damping of the laser cooling $\gamma_\mathrm{L}$ increases $R_\mathrm{L}$ and has the effect of lowering the coupling rate of the Be$^+$ ions to the resonator $\tilde{\gamma}_\mathrm{Be}$, which, in the absence of laser cooling is given by the dip width $\gamma_\mathrm{Be} \propto N_\mathrm{Be}$. For a given number of laser-cooled ions $N_\mathrm{Be}$, $\gamma_\mathrm{L}$ must be optimized and in the limiting case when $\gamma_\mathrm{L} \ll \gamma_\mathrm{Be}$ the Be$^+$ are driven by the resonator and the dip signal is unchanged. Likewise, when $\gamma_\mathrm{L} \gg \gamma_\mathrm{Be}$, the Be$^+$ ions are decoupled from the resonator and the dip signal vanishes. In both limiting cases, the temperature of the resonator and the proton remain unchanged.

However, increasing $N_\mathrm{Be}$ increases $\tilde{\gamma}_\mathrm{Be}$ and laser cooling reduces the temperature of the resonator and the proton, even at large $\gamma_\mathrm{L}$. To lower the temperature and to investigate the scaling of $T_\mathrm{p}$, we performed a series of further measurements with varying $N_\mathrm{Be}$ and laser detuning $\delta$ (Fig. 4). We additionally analysed the temperature scaling by comparing to a temperature model in which the common mode temperature $T_\mathrm{CM}$ of the equivalent circuit arises from competing dissipation sources; the noise temperature of the environment, $T_0$, at a coupling rate given by the width of the LC resonance $\gamma_\mathrm{D}$, and to the Be$^+$ ions at temperature $T_\mathrm{Be}$. As a result, the system comes to thermal equilibrium at

$$T_\mathrm{p} = T_\mathrm{CM} \tag{4}$$

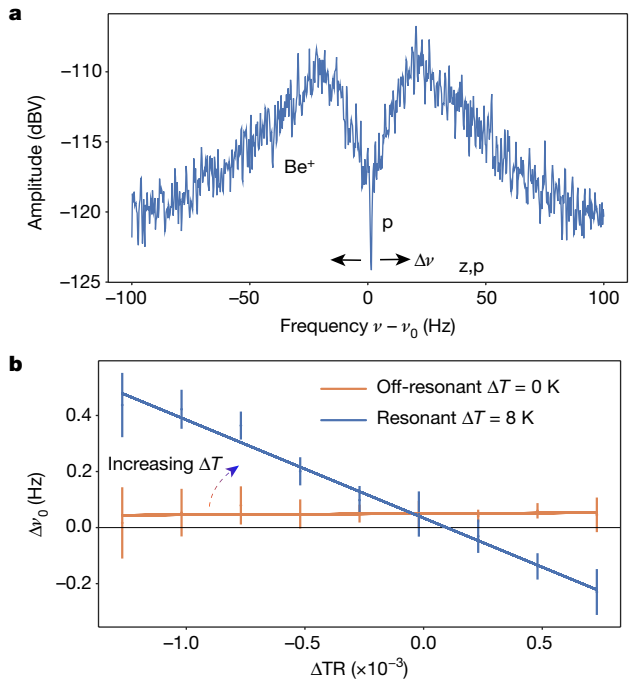

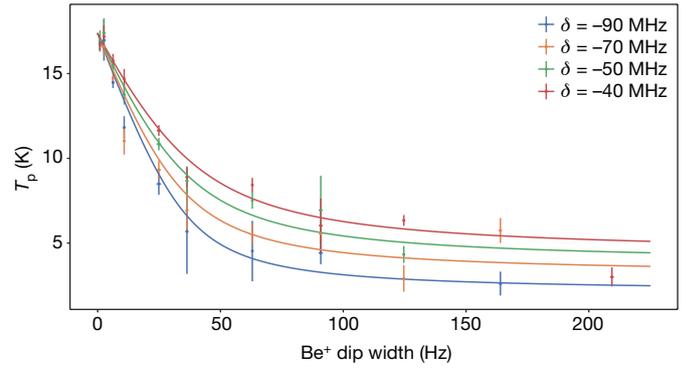

**Fig. 4 | Temperature scaling.** The temperature of the cooled proton-resonator system is shown as a function of the laser detuning $\delta$ and the coupling of the laser-cooled Be$^+$ ions to the resonator parametrized by the dip width. Vertical error bars, fit uncertainty of the measured slope. Smaller horizontal error bars, standard error of $\gamma_{Be}$. Each point consists of a minimum of 50 measured frequency shifts, although more data was collected at some detunings, resulting in lower uncertainties. Fits to the data are shown to illustrate the $1/N_{Be}$ scaling when the dip width becomes larger than the resonator width; approximately 40 Hz.

**Fig. 3 | Sympathetic cooling demonstration. a**, A cloud of laser-cooled Be$^+$ ions appears on the resonator as a broad dip that reduces the temperature of that resonator mode. The proton spectral dip is much narrower and continues to short the resonator noise, and can be used as a temperature probe via equation (3). **b**, The frequency shift $\Delta\nu_z$, from which $\Delta T$ is extracted, is shown for $N_{Be} \approx 10$ and laser detuning $\delta \approx 100$, as a function of the tuning ratio $\Delta TR$, given by $V_{CE}/V_0$. Each data point is an average of several measurements. Error bars, standard deviation of the temperature reduction that arises from the scatter of the final proton temperature. The change in temperature is extracted from the slope, which, when combined with the resonator temperature, allows us to extract the final temperature of the proton.

$$T_{CM} = \frac{T_0 \gamma_D + T_{Be} \tilde{\gamma}_{Be}}{\gamma_D + \tilde{\gamma}_{Be}}. \quad (5)$$

When $\tilde{\gamma}_{Be} \gg \gamma_D$, the proton temperature is approximated as

$$T_p \approx T_0 \frac{\gamma_D}{\tilde{\gamma}_{Be}} + T_{Be}, \quad (6)$$

reproducing the $1/N_{Be}$ scaling, by $\gamma_{Be} \propto N_{Be}$, appearing in the non-resonant proposal[9,11] and related proposals in the context of trapped ion quantum information[10,33,34]. In these measurements, the laser detuning $\delta$ can be viewed as a tuning parameter that changes the $\gamma_L$ and subsequently $\tilde{\gamma}_{Be}$ (see Methods). As a result, the lowest proton temperatures are not found by minimizing the Be$^+$ temperature, which would correspond to lower laser detunings, but by maximizing the coupling of the ions to the detector, corresponding to larger $\tilde{\gamma}_{Be}$. For the experimental parameters used here, $\delta = -90$ MHz is the largest laser detuning at which the proton dip is still visible. The largest ion cloud $\gamma_{Be} = 164(5)$ Hz and largest detuning from the centre of the cooling transition $\delta = -90$ MHz (Fig. 4) is representative of the lowest temperatures observed in our measurements. We achieve a temperature reduction of

$$\Delta T_p = 14.4 \, (0.7) \, K, \quad (7)$$

and using the environment temperature $T_0$, we obtain

$$T_p = T_0 - \Delta T_p = 2.6 \, (2.5) \, K, \quad (8)$$

with uncertainty dominated by the one of $T_0$. This measurement demonstrates a temperature reduction of 85%.

Lower temperatures can be achieved by lowering the noise temperature of the amplifier, $T_0$, increasing the Q-value of the resonator, or by operating with smaller traps that increase $\gamma_{Be}$ quadratically with lower radius. In addition, by performing these demonstration measurements fully on-resonance for maximal coupling rates, the balance of heating by the resonator to cooling by the Be$^+$ ions is maximally inefficient, and future cooling work will be done off-resonantly to balance the coupling rate and the temperature limit with engineered cooling sequences[35].

In the context of our experimental goals, this technique can be readily applied to sympathetically laser cool protons and antiprotons in the same large macroscopic traps that enable precision measurements of the charge-to-mass ratio and g-factor[1,11]. In addition, while we measure the axial temperature, sideband coupling[29] or axialization[36] can be used to cool the radial motion of the antiproton. In measurements of nuclear magnetic moments, this will enable nearly 100% spin-flip fidelity[11,15,16,37], and can reduce the dominant systematic effect proportional to the particle temperature in the highest precision mass measurements[38–40]. In addition, this technique can be used to cool other exotic systems such as highly charged[23,24] or molecular ions[17,41] and the sympathetically cooled resonator can enhance the sensitivity of dark matter searches[13,41,42]. Ultimately, this demonstration realises a long-sought experimental technique that will enable precision experiments of any charged species at lower temperatures.

## Online content

Any methods, additional references, Nature Research reporting summaries, source data, extended data, supplementary information, acknowledgements, peer review information; details of author contributions and competing interests; and statements of data and code availability are available at https://doi.org/10.1038/s41586-021-03784-w.

**BASE Collaboration**

M. Bohman[1,2 ✉], V. Grunhofer[3], C. Smorra[2,3], M. Wiesinger[1,2], C. Will[1], M. J. Borchert[2,4,5], J. A. Devlin[2,6], S. Erlewein[2,6], M. Fleck[2,7], S. Gavranovic[3], J. Harrington[1,2], B. Latacz[2], A. Mooser[1], D. Popper[3], E. Wursten[2,6], K. Blaum[1], Y. Matsuda[7], C. Ospelkaus[4,5], W. Quint[8], J. Walz[3,9] & S. Ulmer[2]

## Methods

### Equations of motion for the coupled ion-trap systems

The axial motions of the trapped proton and Be$^+$ ions are described by a harmonic oscillator driven by the oscillating voltage on the trap electrodes connected to the resonant LC circuit, $V_{LC}$, and in the case of the Be$^+$ ion(s) by an additional photon scattering force from the cooling laser, $F_L$:

$$m_p \ddot{z}_p + m_p \omega_{z,p}^2 z_p = \frac{q}{D_p} V_{LC}$$
$$m_{Be} \ddot{z}_{Be} + m_{Be} \omega_{z,Be}^2 z_{Be} = \frac{q}{D_{Be}} V_{LC} + F_L. \quad (9)$$

$V_{LC}$ is composed of voltage noise from the environment, $V_{noise}$, and the voltage arising from image currents induced by the proton and Be$^+$ ions, $I_p$ and $I_{Be}$, respectively. On resonance $\omega_R = \omega_{z,p} = \omega_{z,Be}$, the impedance of the LC circuit is given by its equivalent parallel resistance, $R_p$, and

$$\frac{1}{\omega_R^2} \ddot{V}_{LC} + \frac{1}{Q \omega_R} \dot{V}_{LC} + V_{LC} = L_R(\dot{I}_{noise} + \dot{I}_p + \dot{I}_{Be}). \quad (10)$$

$I_{noise}$ is the noise current from the environment, and $I_p$ and $I_{Be}$ are the induced image currents of the proton and the Be$^+$ ions, respectively:

$$I_{p,Be} = \frac{q}{D_{p,Be}} \dot{z}_{p,Be}, \quad (11)$$

where $D_{p,Be}$ are the trap-dependent effective electrode distances[28].

The equations of motion and the equation for the voltage in the LC circuit form a set of coupled stochastic differential equations without closed analytical solutions available. As a result, we analyse the frequency response of the system by calculating the impedance of the equivalent circuit in Fig. 1b and estimate the energy of the proton by calculating the temperatures of each component based on their energy exchange rates. Finally, we numerically integrate the differential equations in simulations that allow the comparison of FFT spectra and the visualization of the time-domain behaviour in the system.

### Impedance analysis of the equivalent circuit

The FFT spectrum in Fig. 1c results from the noise on the image-current detector $u_n^2 = 4 k_B T_0 \text{Re}[Z(\omega)] \Delta f$ at effective noise temperature $T_0$, FFT bandwidth $\Delta f$, and the impedance $Z(\omega)$ of the circuit in Fig. 1b. The lineshape of resistively cooled particles stored in a single trap based on the impedance of the equivalent circuit is well understood[28,43]. Here, we evaluate the impedance for two independently biased ion-trap systems as:

$$\frac{\text{Re}[Z(\omega)]}{R_p} = \left(1 + \frac{k_L}{k_L^2 + \delta_{Be}(\omega)^2}\right) \bigg/ \left(1 + \frac{k_L}{k_L^2 + \delta_{Be}(\omega)^2}\right)^2 + \delta_R(\omega)^2 + \delta_p(\omega)^{-2} +$$

$$\frac{\delta_{Be}(\omega)^2}{(k_L^2 + \delta_{Be}(\omega)^2)^2} - \frac{2\delta_R(\omega)}{\delta_p(\omega)} - \frac{2\delta_R(\omega)\delta_{Be}(\omega)}{(k_L^2 + \delta_{Be}(\omega)^2)} + \frac{2\delta_{Be}(\omega)}{\delta_p(\omega)(k_L^2 + \delta_{Be}(\omega)^2)}\bigg), \quad (12)$$

where $k_L = R_L/R_p$ allows for additional damping in one of the traps. The lineshapes of the individual components arise from $\delta_i(\omega) = 2(\omega - \omega_i)/\gamma_i$, which are parameters proportional to the ratio of the frequency detuning ($\omega - \omega_i$) to the oscillator linewidth, $\gamma_i$. The index $i \in \{R, Be, p\}$ relates to the resonator, the Be$^+$ ions and the proton, respectively. In the absence of additional damping $k_L = 0$, the impedance simplifies to

$$\frac{\text{Re}[Z(\omega)]}{R} = \frac{1}{1 + \delta_R^2 + \delta_p^{-2} + \delta_{Be}^{-2} - 2\frac{\delta_R}{\delta_p} - 2\frac{\delta_R}{\delta_{Be}} + 2\frac{1}{\delta_p \delta_{Be}}}, \quad (13)$$

which describes the lineshape of the data shown in Fig. 1c. Similarly, the heat maps in Fig. 2a and Fig. 2b compare the FFT spectra from experiment to ones calculated with $Z(\omega)$, and show consistent behaviour.

With laser cooling included, $R_L > 0$ and the dip feature of the Be$^+$ ions is modified as shown in Fig. 3a. The corresponding impedance is calculated for varying $R_L$ in Extended Data Fig. 1. In both cases, regardless of the value of $R_L$, the proton shorts the noise of the LC circuit on resonance. The Be$^+$ ions decouple from the LC circuit as $R_L$ reduces the fraction of noise power dissipated in the series LC circuit of the Be$^+$ ions—ultimately leading to a vanishing dip signal. This decoupling effect is well known from other coupled oscillator systems[44] and motivates the reduced coupling of the Be$^+$ ions to the LC circuit, $\tilde{\gamma}_{Be} < \gamma_{Be}$.

### Temperature model

The temperature model presented here is described in ref.[45] and assumes that each component of the three-oscillator system consisting of the trapped proton, the trapped Be$^+$ ion(s) and the resonator comes to thermal equilibrium with the rest of the system at temperatures defined by the energy exchange rates in the system.

The Be$^+$ ions are damped by the resonator as well as the cooling laser, and the power transmitted by the Be$^+$ ions is then written

$$\left\langle \frac{dE_{Be}}{dt} \right\rangle = k_B T_{Be} \tilde{\gamma}_{Be} + \left\langle \frac{dE_{Be}}{dt} \right\rangle_{laser}, \quad (14)$$

where $\left\langle \frac{dE_{Be}}{dt} \right\rangle_{laser}$ is the the power dissipated by scattered photons. An identical analysis applies to the resonator, which is coupled to the environment with a coupling rate $\gamma_D$ given by the width of the resonance, or the Q-value, and to the Be$^+$ ions with a coupling rate $\tilde{\gamma}_{Be}$. These relations produce the system of equations shown in the main text,

$$T_p = T_{CM} \quad (15)$$

$$T_{CM} = \frac{T_0 \gamma_D + T_{Be} \tilde{\gamma}_{Be}}{\gamma_D + \tilde{\gamma}_{Be}}. \quad (16)$$

The power dissipated by the resonator while the Be$^+$ ions are laser cooled can be written as

$$k_B T_{CM} \tilde{\gamma}_{Be} = \langle \tilde{J}_z \rangle R_p, \quad (17)$$

and in combination with the power dissipated by the resonator in the absence of laser cooling,

$$k_B T_0 \gamma_{Be} = \langle J_z \rangle R_p, \quad (18)$$

allows the reduced coupling rate to be written as

$$\tilde{\gamma}_{Be} = \frac{T_0}{T_{CM}} k \gamma_{Be}, \quad (19)$$

where $k$ is defined by the ratio

$$k = \frac{\langle \tilde{J}_z^2 \rangle}{\langle J_z^2 \rangle}. \quad (20)$$

Although $k$ can, in principle, be extracted from the FFT spectrum, the extraction of individual $k$ values is imprecise and $k$ and $T_{Be}$ are treated as constant fit parameters in Fig. 4. A more accurate determination of $\left\langle \frac{dE_{Be}}{dt} \right\rangle_{laser}$ can be performed by measuring the photon scattering rate[9] and is planned for future measurements.

## Simulations and time-domain behaviour

We access the time-domain behaviour of the proton–ion–resonator system through simulations, which are performed by numerically integrating equation (9) and equation (10). By replacing $V_{LC} = L_R \dot{i}_L$, where $I_L$ is the current flowing through the inductance $L_R$, these equations can be rewritten as

$$m_p \ddot{z}_p + m_p \omega_{z,p}^2 z_p = \frac{q}{D_p} L_R \dot{i}_L \; m_{Be} \ddot{z}_{Be} + m_{Be} \omega_{z,Be}^2 z_{Be}$$
$$= \frac{q}{D_{Be}} L_R \dot{i}_L + F_L L_R C_R \ddot{i}_L + \frac{L_R}{R_p} \dot{i}_L + I_L + I_{noise} + \frac{q}{D_p} \dot{z}_p + \frac{q}{D_{Be}} \dot{z}_{Be} \quad (21)$$
$$= 0.$$

The integration is performed in time steps of $\Delta t = 1$ ns for most simulations and the equivalent thermal noise $\langle I_{noise}^2 \rangle = 4 k_B T_0 \Delta f / R_p$ is computed in each step $n$ as

$$I_{noise,n} = \sqrt{2 k_B T_0 / (R_p \Delta t)} \cdot G_n(\mu = 0, \sigma = 1), \quad (22)$$

where, owing to the discrete time steps, the noise bandwidth is defined as $\Delta f = 1/(2\Delta t)$. $G_n(\mu = 0, \sigma = 1)$ is a Gaussian distribution with mean $\mu = 0$ and standard deviation $\sigma = 1$ which is sampled every step, conserving the standard deviation of the noise while fulfilling the criterion that two subsequent values must be uncorrelated.

We implement laser cooling in the simulations by assigning a photon absorption probability to an ion in its electronic ground state at each time step. The laser, at wavelength $\lambda$, with wave vector $k_L$ in the axial direction, is detuned from the centre of the transition frequency at $f_0$ by a detuning $\delta$ and we assume that the linewidth of the laser is negligible compared to the transition linewidth $\Gamma$. The discrete photon absorption probability depends on the velocity of the ion due to the Doppler effect and, in the low saturation limit, can be written as

$$P_{abs}(v_z) = \frac{I}{I_{sat}} \frac{1/2}{1 + 4\left(\frac{2\pi\delta + k_L v_z}{\Gamma}\right)^2} \Gamma \Delta t, \quad (23)$$

where the saturation intensity is $I_{sat} = \frac{2\pi^2 \hbar \Gamma}{3\lambda^3}$ and $I$ is a free parameter that is tuned to match the laser intensity in the experiment. Upon absorption of a photon the ion transitions to the excited electronic state and receives a momentum kick $-\hbar k_L$. The ion decays to the ground state via spontaneous emission with probability $\Gamma \Delta t$ and receives a momentum kick in the axial direction of $\hbar k \cos\theta$ where the angle $\theta$ accounts for radial momentum of the emitted photon. The ion can also decay via stimulated emission, in which case the momentum kick is $+\hbar k_L$.

Data preparation and analysis are performed in R[46], while the intensive part of the calculation is performed using C++ via the Rcpp-package[47]. We use a fourth order symplectic integrator[48] to calculate the particle trajectories and the voltage across the RLC circuit to ensure that energy is, on average, conserved for numeric integration with more than $10^{10}$ steps.

In simulations of the p-Be$^+$ resonator system, we apply the conditions of the experiments described in the main text to reproduce the frequency-domain behaviour in Extended Data Fig. 2, with Figs. 2a, b corresponding to the experimental results shown in the inset of Figs. 2d and 3a), respectively. The evolution of the oscillator energies with ten Be$^+$ ions, $N_{Be} = 10$, is shown in Extended Data Fig. 2c. Starting from $t = 7.5$ s, a parametric drive is applied, resulting in a significant increase in the energy of the proton and the Be$^+$ ions from an initial temperature of 17 K. Similarly, Extended Data Fig. 2d shows the energy exchange between a single proton, 80 Be$^+$ ions, and the resonator all on resonance, where the cooling laser is applied from $t = 10$ s, resulting in rapid cooling of the Be$^+$ ions and a temperature reduction of the proton.

## Axial frequency shifts

A particle in a Penning trap is subjected to shifts in the mode frequencies due to the inhomogeneity of the magnetic field and the anharmonic contributions to trapping potential[26,32]. The magnetic field in the trap centre can be written with the lowest order corrections as

$$\mathbf{B}(z, r) = B_0 \hat{z} + B_2((z^2 - r^2/2)\hat{z} - rz\hat{r}), \quad (24)$$

where a quadratic gradient $B_2$ shifts the axial frequency as a function of the radial energy by

$$\Delta v_z = \frac{1}{4\pi^2 m v_z} \frac{B_2}{B_0} E_+. \quad (25)$$

We use this effect to demonstrate the energy exchange between the heated Be$^+$ ions and the proton in Fig. 2c. Here, the proton axial mode and modified cyclotron mode are sideband-coupled with a quadrupolar rf drive, so that after the sideband coupling, the proton cyclotron energy freezes out at an energy $E_+ = (v_+/v_z)E_z$, where $E_z$ is the axial energy while coupling the axial mode to the excited Be$^+$ ions.

timized to have a homogeneous magnetic field and is unsuited for energy measurements using equation (25) at low energy, with a temperature resolution of <0.1 mHz K$^{-1}$. In the the sympathetic cooling measurements presented here, we instead used the trapping potential anharmoncity that we introduced in the proton trap to determine the temperature of the trapped particle. The trapping potential can be expanded in terms of $C_n$ coefficients[26,31,32], and the higher-order terms $C_{2n}$, $n \geq 2$ shift the trap frequencies $v_i$ by

$$\Delta v_i \propto C_{2n} E_j^{n-1}, \quad (26)$$

where $E_j$ is the energy of a trap mode. The coefficients $C_n$ can be written in terms of a 'tuning ratio', TR, defined by the ratio of the voltage applied to the central ring electrode to the voltage applied to a correction electrode, as

$$C_n = E_n + D_n \text{TR}. \quad (27)$$

$D_n$ can be calculated from the trap geometry, and the axial frequency shift due to the leading energy-dependent trap anharmonicity $C_4$ can be written as

$$\Delta v_z = \frac{1}{4\pi^2 m v_z} \frac{3}{2} \frac{D_4}{C_2} k_B T_z \Delta \text{TR}, \quad (28)$$

where $\Delta$TR is the offset in applied tuning ratio from the ideal tuning ratio at which $C_4 = 0$. Ultimately, the axial frequency shift as a function of TR and the axial energy $E_z$ can be expressed as

$$\Delta v_z = \kappa_{D_4} E_z, \quad (29)$$

where, for a proton stored in the proton trap, $\kappa_{D_4} = 45.4 \Delta$TR Hz K$^{-1}$. This effect is used to determine the change of the proton axial temperature while the resonator is cooled with the laser-cooled Be$^+$ ions, and is the underlying method for the data shown in Figs. 3 and 4.

Temperature measurements using this method are limited by the determination of $T_0$ to the ~1 K level. We have previously performed higher precision temperature measurements using a dedicated, spatially distant trap with a ferromagnetic ring electrode that uses the shift of equation (25) to obtain a cyclotron energy resolution of up to 80 Hz K$^{-1}$ (refs. [15,49]) and have developed a similar trap to reach 10 mK temperature resolution in future cooling measurements.

We also note that equation (28) and equation (29) cause the axial dip to spread out during an FFT averaging window and decrease the dip signal-to-noise ratio. This is reflected in the increased uncertainties at larger $\Delta$TR that can be seen in Fig. 3b.

## Parasitic drive heating

The demonstration of remote energy exchange was performed by exciting a small cloud of Be$^+$ with an rf drive at twice the resonance frequency, $2\nu_0$, with results presented in Fig. 2d. To confirm that the proton is excited only by the resonantly coupled ions, we performed a series of background measurements. These control measurements show that the proton is excited only when resonant with the Be$^+$ ions and that the proton is unaffected by the excitation drive when the Be$^+$ ions are detuned (Extended Data Fig. 3).

Although the measurements presented in Fig. 2d are performed by interleaving the on- and off-resonant configurations, we additionally analysed the temperature of the proton in the presence of the drive. During these measurements, we transferred the axial energy to the modified cyclotron mode and measured the resulting energy dependent axial frequency shift, described in the main text. We measured an axial temperature of

$$T_{z,\text{on}} = 32(2) \text{ K} \tag{30}$$

in the presence of the drive, and an axial temperature of

$$T_{z,\text{off}} = 20(5) \text{ K} \tag{31}$$

in the absence of the drive, where the error comes from the fit uncertainty of the frequency scatter distribution. As a result, we constrain the possible increase in axial temperature due to the excitation drive to no more than a factor of two. From the spectra shown in Extended Data Fig. 3, we see that the signal-to-noise ratio of the proton dip is unaffected by the drive and conclude that an increase in axial temperature would come not from direct coupling of the excitation drive to the proton but via an increase in the equivalent noise temperature of the resonator. We further note that the resonator temperature of 17.0 K given in the main text comes from the weighted mean of several temperature measurements performed with several methods. Importantly, residual heating due to the drive is far lower than the energy scatter shown in Fig. 2d; approximately 40,000 K $k_B^{-1}$.

## Data availability

The datasets generated and/or analysed during this study are available from the corresponding authors on request. Source data are provided with this paper.

## Code availability

The code used during this study is available from the corresponding authors on reasonable request.

**Acknowledgements** This study comprises parts of the PhD thesis work of M.B. We acknowledge the contributions of G. L. Schneider and N. Schön to the design and construction of the Penning-trap system. We thank S. Sturm for helpful discussions regarding the cooling method presented here and acknowledge similar developments toward cooling highly charged ions in the ALPHATRAP collaboration. We acknowledge financial support from the RIKEN Chief Scientist Program, RIKEN Pioneering Project Funding, the RIKEN JRA Program, the Max-Planck Society, the Helmholtz-Gemeinschaft, the DFG through SFB 1227 'DQ-mat', the European Union (Marie Skłodowska-Curie grant agreement number 721559), the European Research Council under the European Union's Horizon 2020 research and innovation programme (Grant agreement numbers 832848-FunI and 852818-STEP) and the Max-Planck-RIKEN-PTB Center for Time, Constants and Fundamental Symmetries.

**Author contributions** M.B., A.M., S.U., J.W. and M.W. designed the experimental apparatus. M.B., A.M. and M.W. assembled the trap and laser systems. M.B., V.G., C.S. and M.W. contributed to the experiment run. M.B. and C.S. implemented the methods, and recorded and evaluated the experimental data. C.W. developed the simulation code and provided the simulation results. M.B., C.S., K.B. and S.U. prepared the manuscript, which was discussed and approved by all authors.

**Funding** Open access funding provided by Max Planck Society.

**Competing interests** The authors declare no competing interests.

**Additional information**
**Supplementary information** The online version contains supplementary material available at https://doi.org/10.1038/s41586-021-03784-w.
**Correspondence and requests for materials** should be addressed to M.B.
**Peer review information** *Nature* thanks Manas Mukherjee, Richard Thompson and the other, anonymous, reviewer(s) for their contribution to the peer review of this work.
**Reprints and permissions information** is available at http://www.nature.com/reprints.

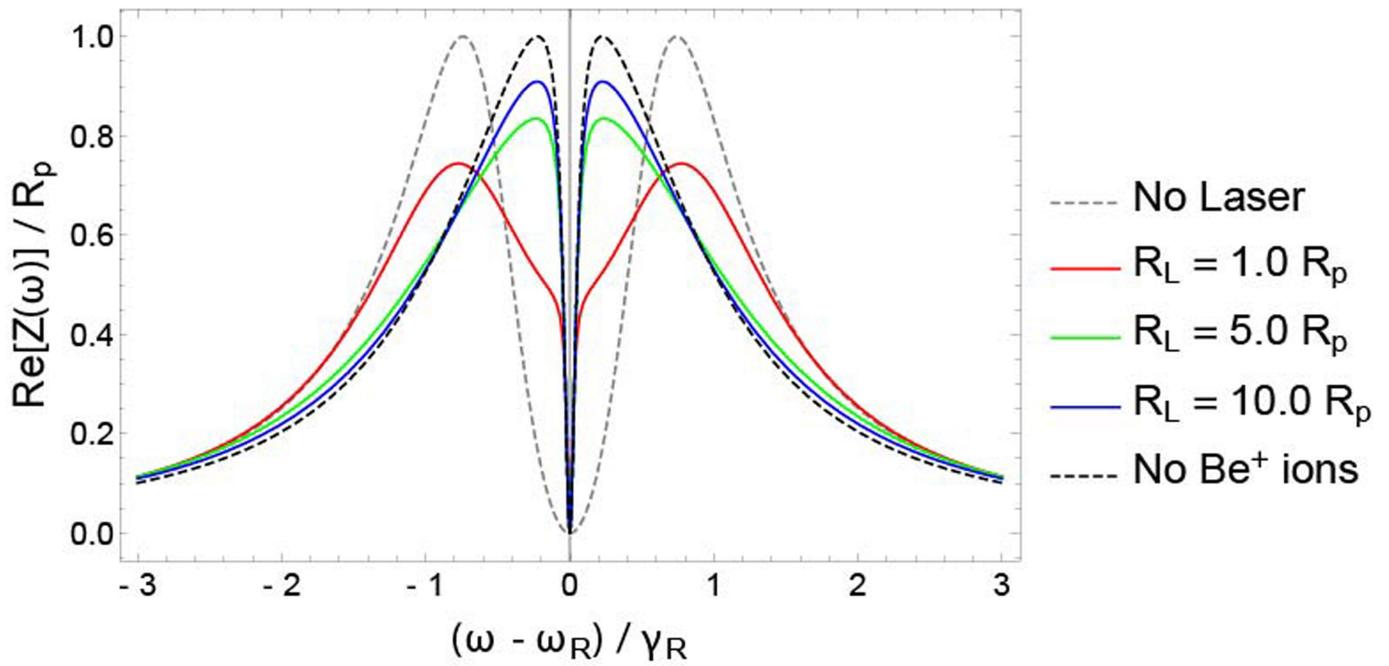

**Extended Data Fig. 1 | Calculated impedance of the equivalent circuit.** The normalised impedance $Re[Z(\omega)]/R_p$ is plotted for different values of $R_L$, where no damping corresponds to $R_L = 0$, and no Be$^+$ ions to $R_L \to \infty$. Here, $\gamma_R = 2\gamma_{Be} = 20\gamma_p$, corresponding in experiment to about 30 Be$^+$ ions and a single proton.

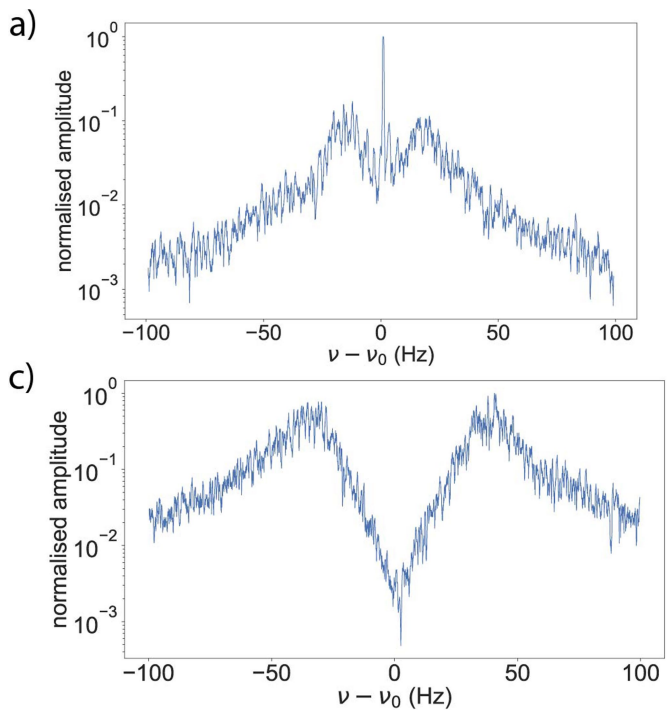 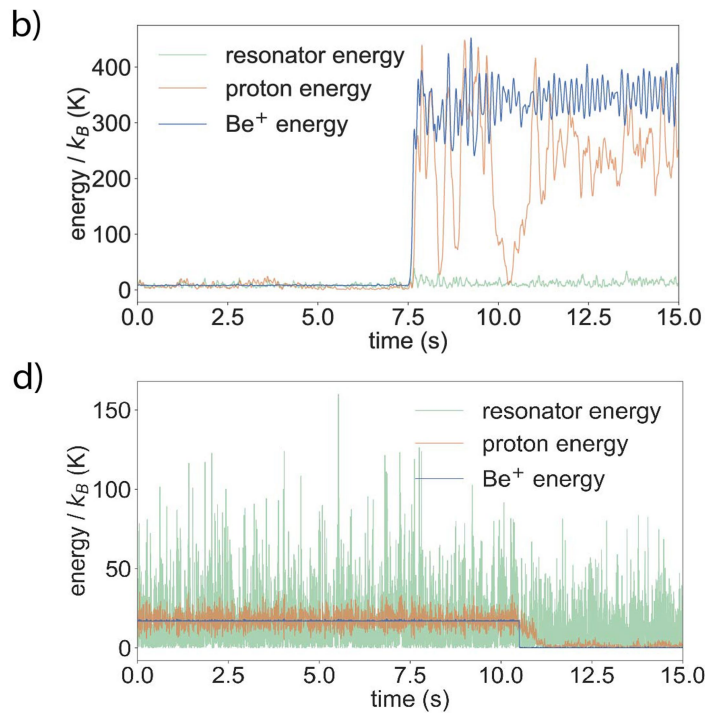

**Extended Data Fig. 2 | Simulation results. a)** A computed FFT spectrum is shown simulating the experimental conditions of Fig. 2d) in the main text. **b)** Representative time domain behaviour for these measurements is shown where the excitation drive is applied at time at time $t = 10$s. **c)** A computed FFT spectrum is shown simulating the experimental conditions of Fig. 3a) in the main text. **d)** Representative time domain behaviour for these measurements is shown where the cooling laser is applied at time $t = 10$s.

# Article

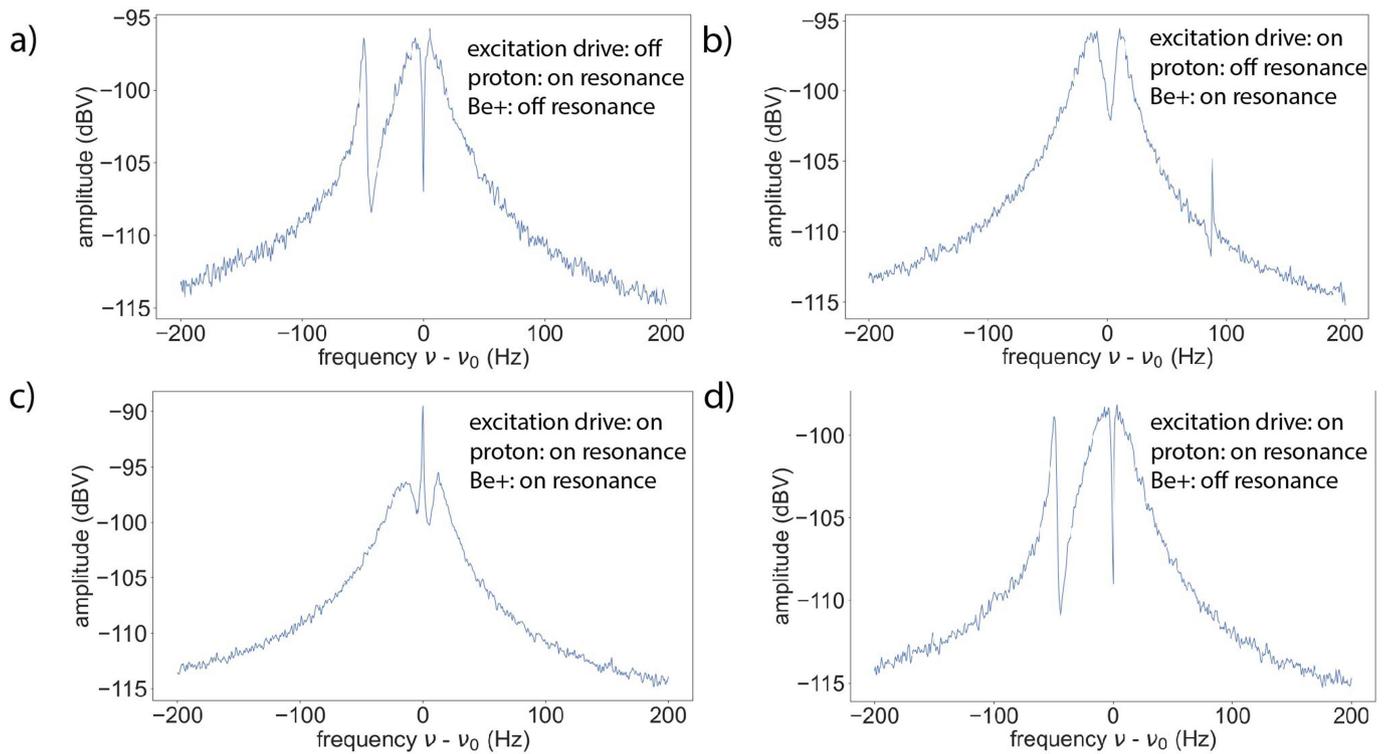

**Extended Data Fig. 3 | Excitation drive background measurements. a)** An FFT spectrum while the excitation drive is off, the proton is on resonance with the resonator and the Be$^+$ ions are off resonance. **b)** An FFT spectrum while the excitation drive is on, the proton is off resonance with the resonator and the Be$^+$ ions are on resonance. **c)** An FFT spectrum while the excitation drive is on, the proton is on resonance with the resonator and the Be$^+$ ions are on resonance. **d)** An FFT spectrum while the excitation drive is on, the proton is on resonance with the resonator and the Be$^+$ ions are off resonance.